\documentstyle[12pt,fleqn]{article}
\def\singlespace {\smallskipamount=3.75pt plus1pt minus1pt
                  \medskipamount=7.5pt plus2pt minus2pt
                  \bigskipamount=15pt plus4pt minus4pt
                  \normalbaselineskip=15pt plus0pt minus0pt
                  \normallineskip=1pt
                  \normallineskiplimit=0pt
                  \jot=3.75pt
                  {\def\smallskip {\vskip\smallskipamount}}
                  {\def\medskip   {\vskip\medskipamount}}
                  {\def\bigskip   {\vskip\bigskipamount}}
                  {\setbox\strutbox=\hbox{\vrule 
                    height10.5pt depth4.5pt width 0pt}}
                  \parskip 7.5pt
                  \normalbaselines}

\begin{document}

\singlespace
\rightline {TIFR-TAP Preprint}
\begin{center}
{\Large {Negative Pressure and Naked Singularities}}\\
{\Large {in Spherical Gravitational Collapse}}
\end{center}
\vspace{1.0in}
\vspace{12pt}
\begin{center}
{\large{F. I. Cooperstock\footnote{Permanent address: Department of Physics 
and Astronomy, University of Victoria, Victoria, B. C., Canada V8W 3P6}, 
S. Jhingan, P. S. Joshi and T. P. Singh\\
Theoretical Astrophysics Group\\
Tata Institute of Fundamental Research\\
Homi Bhabha Road, Bombay 400005, India.\\
}}
\end{center}
\vskip 1 in
\centerline{\bf ABSTRACT}

\medskip

\noindent Assuming the weak energy condition, we study the nature of the 
non-central shell-focussing singularity which can form in the 
gravitational collapse of a spherical compact object in classical general
relativity. We show that if the radial pressure is positive, the singularity
is covered by a horizon. For negative radial pressures, the singularity will 
be covered if the ratio of pressure to the density is greater than $-1/3$ 
and naked if this ratio is $\leq \ -1/3$. 

\bigskip

\noindent PACS Nos. 04.20.-q, 04.20D, 04.70, 97.60.L

\vfil\eject

There are various known examples of formation of black-holes and
naked singularities in spherical gravitational collapse (for recent
reviews see \cite{reviews}). However the general conditions on the
initial data in spherical collapse which will ensure formation of a 
black-hole or of a naked singularity are not well-understood. In this Letter
we demonstrate the role of a negative radial pressure in the formation of
a naked singularity in the gravitational collapse of a spherical compact
object. We use methods similar to those developed by Podurets \cite{Podurets}
and Lifshitz and Khalatnikov \cite{LK} to describe solutions of Einstein's
equations in the vicinity of a singularity.

In comoving coordinates $(t,r,\theta,\phi)$ the spherically symmetric
metric is
\begin{equation}
   ds^{2} = e^{\sigma}\;dt^{2}  -  e^{\omega}\;dr^{2} - R^{2} \; d\Omega^{2}
   \label{metric}
\end{equation}
where $\sigma$, $\omega$ and $R$ are functions of $t$ and $r$ and $R(t,r)$
is the area radius at time $t$ of the fluid element with comoving label $r$.
The energy-momentum tensor is $T_{ik}=diag(-\rho, p_{r}, p_{\theta}, 
p_{\theta})$ where $p_{r}$ and $p_{\theta}$ are respectively the radial and
tangential pressures. We assume the weak energy condition: $\rho\geq 0$,
$\rho + p_{r} \geq 0$ and $\rho + p_{\theta} \geq 0$. The field equations are
(see e.g. \cite{DJ94})
\begin{equation}
    m'=4\pi\rho R^{2}R'\;,   \label{mprime}
\end{equation} 
\begin{equation}
    \dot{m} = -4\pi p_{r}R^{2}\dot{R}\;,   \label{mdot}
\end{equation}
\begin{equation}
   2p_{\theta}R\dot{R}R' = -2\dot{m}' + m'{\dot{G}\over G}
   + \dot{m} {H'\over H}\;,  \label{rprime}
\end{equation}
\begin{equation}
 -2\dot{R}' +  R'{\dot{G}\over G} + \dot{R} {H'\over H} = 0,\label{rdotprime}
\end{equation}
\begin{equation}
    m = {1\over 2} R \left( 1 + e^{-\sigma}\dot{R}^{2}
         - e^{-\omega}R'^{2}\right) .   \label{energy}
\end{equation}

Prime and dot stand for $\partial/\partial r$ and $\partial/\partial t$
respectively. We have defined $G\equiv e^{-\omega}R'^{2}$ and 
$H\equiv e^{-\sigma}\dot{R}^{2}$. $m(t,r)$ is an arbitrary function 
of integration, and from (\ref{mprime}) it follows, for a fixed time $t$,
that
\begin{equation}
   m(t,r) = 4\pi\int \rho R^{2}\; dR \ + \ m(t,0). \label{mass}
\end{equation}
$m(t,r)$ is interpreted as the mass interior to $r$ at a 
given time $t$. We assume that $m(t,0)=0$ and that the total mass 
$m(t,r_{0})$ is finite, where $r_{0}$ is the boundary of the collapsing
object. The weak energy condition ensures that $m\geq 0$. At the initial 
epoch $t=t_{0}$ we assume $\rho(t_{0},r)$ to be finite and strictly positive
for $r\leq r_{0}$ and zero for $r>r_{0}$. 
Eqn. (\ref{mass}) along with $m(t_{0},0)=0$ then implies $m(t_{0},r)>0$
for $r>0$.

We assume that the gravitational collapse of the star leads to the formation
of a shell-focussing curvature singularity. The singularity curve is given
by $R(t_{s}(r), r)=0$, i.e. the shell with label $r$ shrinks to zero 
area radius at comoving time $t_{s}(r)$. Further, we make the natural
assumption that $(dt_{s}(r)/dR) \neq 0$ $-$ different shells become singular
at different comoving times $t$. Since we are considering collapse,
we take $\dot{R}\leq 0$, the equality sign holding, at best, initially and
in the limit of approach to the singularity. The shell-focussing singularity
at $r=0$ is called the central singularity, and that at $r>0$ non-central.
In this Letter we are concerned with the non-central singularity.  

Assume first that the radial pressure $p_{r}$ remains non-negative throughout
the star during evolution. It then follows from (\ref{mdot}) and the initial
conditions that $m(t, r)$ is non-decreasing. For $r>0$, $m(t,r)$ is strictly 
positive. A shell labelled $r$, with $r>0$, becomes trapped at a time 
$t_{T}(r)$ given by $R(t_{T}(r), r) = 2m(t_{T}(r), r)$. Positivity of $m$ 
implies that $t_{T}(r)<t_{s}(r)$. The shell gets trapped before becoming 
singular and hence the singularity is covered. This argument does not hold
for the $r=0$ singularity, since $m(t,0)=0$. (In that case the ratio $2m/R$
in the limit $R\rightarrow 0, r\rightarrow 0$ will play a decisive role and in
principle the central singularity could be naked.)

Thus the non-central singularity will necessarily be covered if the radial
pressure is positive, irrespective of the sign of the tangential pressure. In 
particular, this includes the case of the radiative equation of state
$p_{r}=p_{\theta}=\rho/3$ which is expected to hold during the late stages of 
collapse. The weak energy condition, however, does allow for $p_{r}$ to be 
negative, in which case it follows from (\ref{mdot}) that in the limit of 
approach to the singularity, $m(t,r)$ for $r>0$ could go to zero and hence the
non-central singularity could in principle be naked, unlike when $p_{r}$
is positive. This also means that any naked singularity arising in
spherical collapse is necessarily massless, in the sense that the function
$m(t,r)$ will go to zero at the naked singularity.

In order to illustrate how a negative radial pressure might alter the nature
of the non-central singularity, we take the example of a perfect fluid with
an equation of state $p_{r}=p_{\theta}\equiv p=k\rho$, with $-1\leq k <0$. 
We could assume that such a distribution might arise in the inner regions of
a star, and is suitably matched to outer regions having positive pressures.
Here we study only the implication of negative pressure, without considering
how such a matching could be achieved. For a perfect
fluid the field equations (\ref{rprime}) and (\ref{rdotprime}) can be written
as
\begin{equation}
       \sigma' = -{2p'\over p + \rho}\;,  \label{sigmaprime}
\end{equation}
\begin{equation}
 \dot{\omega}   =   -{2\dot{\rho}\over p + \rho} - {4\dot{R}\over R}.
                              \label{omegadot}
\end{equation}
Eqns. (\ref{mprime}), (\ref{mdot}) and (\ref{energy}) remain as before,
with $p_{r}=p$. (\ref{sigmaprime}) and (\ref{omegadot}) can be solved
to yield $e^{-\sigma}=\rho^{2k/1+k}$ and $e^{-\omega}=R^{4}\rho^{2/1+k}$,
where we have used coordinate freedom to set constants of integration to
unity. 

Next we change from variables $(t,r)$ to variables $(R,r)$, and the
three basic equations (\ref{mprime}), (\ref{mdot}) and (\ref{energy})
which have to be analysed further become
\begin{equation}
   {\partial m\over \partial R}  =  - 4\pi k\rho R^{2}\;, \label{mdot2}
\end{equation}
\begin{equation}
   {\partial t\over \partial R} {\partial m\over \partial r}  =
   -4\pi (1+k) R^{2} \rho {\partial t\over \partial r}\;, \label{mprime2}
\end{equation}
\begin{equation}
m= {1\over 2} R  +   {1\over 2}R\rho^{2k/1+k}
   {1\over (\partial t/\partial R)^{2}}
   - {1\over 2}R^{5}\rho^{2/1+k}{(\partial t/\partial r)^{2}\over 
     (\partial t/\partial R)^{2}}  .  \label{basic}
\end{equation}   

Following Podurets \cite{Podurets} we make a crucial assumption, namely that
the non-analytic part in $\rho(R,r)$ depends only on $R$ $-$ this is
physically
motivated because there are no special values of $r$. The only exception is
$r=0$ where this assumption is unlikely to hold; in particular it 
does not hold at $r=0$ for Tolman-Bondi dust collapse. Hence the following
analysis is valid only for the non-central singularity, subject to the 
previous analyticity assumption. Thus, $\rho=\psi(R)\rho_{1}(R,r)$ where
$\rho_{1}$ is analytic. By expanding $\rho_{1}(R,r)$ in a power series near
$R=0$ and keeping only the leading term we can approximately express
$\rho$ as
\begin{equation}
     \rho(R,r) = \psi(R) a(r).   \label{rholead}
\end{equation}
We assume that near the singularity curve we can write 
$dt/dr \approx dt_{s}(r)/dr$. Eqn. (\ref{mdot2}) then gives
\begin{equation}
   m(R,r) = -4\pi k a(r) \xi(R)   +  m_{0}(r)    \label{msoln}
\end{equation}
where $d\xi/dR = \psi(R) R^{2}$. Eqn. (\ref{mprime2}) can be written as
\begin{equation}
   {\partial t\over \partial R}  =  -4\pi (1+k) {dt_{s}\over dr}
   a(r)  {\psi(R) R^{2}  \over  (dm_{0}/dr) - 4\pi k \xi(R)(da/dr)}.
   \label{dtdr}
\end{equation}

We first assume that as the singularity is approached, the mass function
$m(R,r)$ goes to a non-zero value. Then in (\ref{basic}) the second term
should either be of the same order as the third term or dominate over it. 
Further, the second term should go to a non-zero value in this limit. 
It can be shown that if these two terms are of the same order then Eqns.
(\ref{basic}), (\ref{msoln}) and (\ref{dtdr}) are inconsistent with each
other. Hence the second term is non-zero in the limit, while the third
term goes to zero. It can then be shown from (\ref{basic}) and (\ref{msoln})
that a necessary condition for the mass to go to a non-zero limit is that
$k\; > \; -1/3$. If $k\; \leq \; -1/3$ the mass goes to zero in the limit.
Note that for $k\; > \; -1/3$ the mass may or may not go to zero in the
limit. If it goes to a non-zero limit, the singularity will be covered,
as discussed above.

Consider next the case that the mass function does go to zero in the limit
of approach to the singularity, and does so as $m(R,r)\sim R^{n}$. From
(\ref{mdot2}) we get $\rho \sim R^{n-3}$, and from (\ref{mprime2}) that
$(\partial t/\partial R) \sim R^{-1}$. Then (\ref{basic}) becomes
\begin{equation}
   m  =  {1\over 2} R   +  A_{0}(r) R^{3 + 2k(n-3)/(1+k)}
         -  B_{0}(r) R^{7 + 2(n-3)/(1+k)},       \label{basic2}
\end{equation}
where $A_{0}(r)$ and $B_{0}(r)$ are positive quantities. 
In order for the density to blow up at the singularity, it is necessary
that $n < 3$. This, along with the original assumption $-1 \leq k < 0$
implies that as $R\rightarrow 0$, the second term in (\ref{basic2}) is
negligible compared to the first one, and hence in this limit (\ref{basic2})
can be written as
\begin{equation}
   m  =  {1\over 2} R  -  B_{0}(r) R^{7 + 2(n-3)/(1+k)}   .    \label{basic3}
\end{equation}

If in the limit $R\rightarrow 0$ we have $m=R/2$, this is precisely the
condition for formation of a trapped surface, and in this case we do not
consider the singularity to be naked. It is at best marginally naked, in
the sense that only one light ray escapes from the singularity. For the
singularity to be naked, it is necessary and sufficient that in the limit
$R\rightarrow 0$, $(2m(R,r)/R) < 1$. This requires that in (\ref{basic3})
the power of the second term should be equal to one, i.e. $n=-3k$. Further,
for nakedness it is necessary that $n\geq 1$ (otherwise $2m/R$ will be
infinite in the limit $R\rightarrow 0$). This gives $k\; \leq \; -1/3$ as
a necessary condition for nakedness. Hence for $k\; > -1/3$ the singularity
will be covered, irrespective of whether or not the mass goes to zero.

For $k \; \leq \; -1/3$ the singularity is naked. This follows because if
it were covered, then the power of the second term in (\ref{basic3}) will be
greater than one, giving $n=1$ and hence $k \; > \; -1/3$. Hence we have the 
final result that for $k \; \leq \; -1/3$ the non-central singularity is 
naked, and for $k \; > \; -1/3$ it is covered. The result is subject to the
crucial but plausible assumption about the analyticity properties of 
$\rho(R,r)$ stated above. The role of negative pressure in naked singularity
formation has been highlighted also in \cite{Szekeres}, although the
starting point and some of the conclusions of that work are different from
ours.

It may appear that the condition $2m/R < 1$ need not be sufficient
for naked singularity formation.  For instance, it was suggested in  
\cite{Shapiro} that the absence of an apparent horizon up until the time
of singularity formation indicates the singularity is naked.
However, it was pointed out in \cite{Wald} and \cite{Jhingan} that although 
one may not detect the apparent horizon at the time of singularity formation, 
in a particular space-time slicing, a trapped surface may nonetheless exist.
This argument of \cite{Wald} and \cite{Jhingan}, however, is not applicable 
to our criterion above, because we are using the area radius $R$ itself as 
one of the coordinates. Hence the result $2m/R < 1$ is independent of the
slicing and indicates the absence of a trapped surface. Of course, the most 
direct way of showing the singularity to be naked is the existence of 
outgoing goeodesics. Nonetheless, our present criterion, though less direct, 
appears to be on the same footing.  

One could ask whether a curvature singularity forms in collapse in spite
of the negative pressure. For instance, it is well-known that in 
Robertson-Walker cosmology the evolution of the scale factor is given by
\begin{equation}
    \ddot{S}(t)  =  -{4\pi\over 3} \left( \rho + 3p \right)S(t).
    \label{ddot}
\end{equation} 
There is a past singularity $S=0$ if $\ddot{S} \leq 0$, i.e. 
$\rho + 3p \geq 0$, 
(i.e. $k \geq -1/3$ for $p=k\rho$).  However a singularity may not form if
$\rho + 3p <0$, (i.e. $k < -1/3$). The weak energy condition ($p > -\rho$)
by itself does not guarantee singularity formation in this case. However,
consider the corresponding equation for $\ddot{R}(t,r)$ in inhomogeneous
collapse, which can be derived by differentiating (\ref{energy}) w.r.t.
$t$ and substituting from (\ref{mdot}) and (\ref{sigmaprime}), 
(see also \cite{Misner}). 
\begin{equation}
    \ddot{R}(t,r) =  -{4\pi\over 3} R e^{\sigma} 
\left( 3p  + {\int 4\pi\rho R^{2}R'dr\over 4\pi R^{3}/3}\right)
  +  {1\over 2}\dot{\sigma}\dot{R}  
  -  {p'Ge^{\sigma}\over R'(p+\rho)}.    \label{rddot}
\end{equation}

(\ref{ddot}) can be recovered from (\ref{rddot}) by setting $R(t,r)=rS(t)$,
$e^{\sigma}=1$ and $G(t,r)=1-kr^{2}$. (\ref{rddot}) differs from (\ref{ddot})
in interesting ways, which suggest that a singularity could form even if
$\rho + 3p <0$. For densities decreasing outwards, the second term inside
the bracket in (\ref{rddot}) is larger than $\rho$, allowing $p$ to be less
than $-\rho/3$ before $\ddot{R}$ becomes positive. Similarly, the second 
and third terms in (\ref{ddot}) are negative in the present case, again
supporting singularity formation by keeping $\ddot{R}$ negative. It is
hence possible that a singularity can form for $k\leq -1/3$, in which case
it will be naked. On the other hand, if a singularity forms only for
$k > -1/3$ it will be covered. We note that there is actually a singularity
theorem \cite{Penrose} which assumes the {\it weak} energy condition.
Lastly we mention that if our assumption of positivity of energy density is
dropped, the mass function could in principle go to a negative value at the
singularity. A negative mass region could arise for instance in the case
of charged sphere solutions of the Einstein-Maxwell theory, leading to
the Reissner-Nordstrom repulsion phenomenon \cite{Rosen}.

We would like to thank Louis Witten for helpful discussions.

\end{document}